\documentclass[12pt]{article}
\usepackage{amsmath,amssymb,pxfonts,graphicx,color}

\newtheorem{prop}{Proposition}

\newtheorem{cor}{Corollary}
\title{Hedging under rough volatility}
\author{Masaaki Fukasawa \\ {\small Osaka University} \and Blanka Horvath \\
  {\small Technical University of Munich \& King's College London} \and
  Peter Tankov\\ {\small ENSAE, Institut
  Polytechnique de Paris}}

\date{}

\begin{document}
\maketitle
\begin{abstract}
  In this chapter we first briefly review the existing approaches to hedging
  in rough volatility models.  Next, we present a simple but general result which
  shows that in a one-factor rough stochastic volatility model, any
  option may be perfectly hedged with a dynamic portfolio containing
  the underlying and one other asset such as a variance swap. In the
  final section we report the results of a back-test experiment using
  real data, where
  VIX options are hedged with a forward variance swap.  In this experiment, using a rough volatility model allows to almost completely
  remove the bias and reduce the overall hedging error by a factor of
  27\% compared to traditional diffusion-based models. 
\end{abstract}


\section{Introduction}
The many advantages of rough volatility models have been outlined in previous chapters. One of the only potential  challenges that remain to be addressed in practice is an (apparent) difficulty to hedge derivatives in rough models.
Hedging in rough volatility models can seem intricate since the dynamics of rough volatility models involve a fractional Brownian motion. In this chapter we demonstrate how this apparent challenge can be overcome in different modelling scenarios exhibiting different levels of generality, which allow us to derive (often explicit) hedging strategies. For building a hedging portfolio, one essentially needs to compute conditional expectations of the form 
\begin{align}\label{eq:Conditional}
C_t=\mathbb{E}[f(S_T)|\mathcal{F}_t],
\end{align} where $f$ is a deterministic payoff function, and
determine the associated martingale representations. 
Classical theory tells us that the option payoff can be replicated at time $t$ for a price $\textbf{P}_{T-t}f(S_t)$, where, $(\textbf{P}_t)_{t\geq 0}$   is the semigroup on $C^b(\mathbb{R}^2)$ generated by the infinitesimal generator $\textbf{A}$ associated with the instantaneous covariance of $S$ and with the (local) martingale problem characterising the law of the process $S$.
Under the assumption that an equivalent local martingale measure exists (and even beyond that, see \cite{ruf2013hedging}), classical theory (\cite{ansel1994couverture,delbaen1995no,jacka1992martingale}) gives conditions when a contingent claim can be hedged (optimally). 
In a classical Markovian setting, 
an optimal trading strategy $h$ can be derived directly 
from 
 the solution $C_t$ to the Cauchy problem $\partial_t C_t(s) -\textbf{A}C_t(s)=0$ associated with the generator \textbf{A}  cf. \cite{ruf2013hedging}. The non-Markovian nature (at least in the finite-dimensional sense) of rough models coming from the fractional Brownian driver leaves most of these results out of scope for rough volatility.
In addition, this very (non-Markovian) nature of rough volatility models
also prohibits the direct use of (PDE-based) efficient numerical methods
for a tractable evaluation of prices \eqref{eq:Conditional} and
associated hedging portfolios, which can make hedging more challenging
than in classical models. In the case of affine rough models it is
possible to exploit the affine structure to derive efficient pricing and
hedging. For the general case Monte Carlo methods have been derived
\cite{bennedsen2017hybrid, fukasawa2021refinement, mccrickerd2018turbocharging} for pricing under rough volatility\footnote{For small values of the Hurst parameter the computational cost of calculation of prices increases.} which can in some cases be computationally slow, but by using deep neural networks it is possible to speed up these pricing methods by several orders of magnitude as demonstrated in \cite{bayerdeep,horvath2021deep}.  
Also, deep neural networks can aid the direct computation of hedging
strategies as in \cite{buehler2019deep,horvath2021bdeep}. Indeed, the deep hedging
framework is applicable in great generality, including rough
volatility models as recently demonstrated in \cite{horvath2021bdeep}. These
hedging portfolios are obtained, based on the idea that every
investment strategy gives rise to a profit and loss, whose
distribution can be optimised with respect to specific risk
measures. More specifically, Horvath Teichmann and Zuric \cite{horvath2021bdeep}
compute hedging strategies for the rough Bergomi model, numerically
building on results of Gassiat \cite{gassiat2019martingale} and Viens and Zhang
\cite{viens2019martingale} and demonstrate the applicability of deep hedging for
the calculation of hedging strategies in rough models. \\

\paragraph{The special role of the forward variance curve in rough
    models} As mentioned, in some cases within the rough volatility
    framework, it is possible to derive hedging strategies more
    explicitly. In the rough Heston model for example, El Euch and Rosenbaum~\cite{el2018perfect} obtain explicit hedging strategies that lead to \emph{perfect hedging}.
For obtaining these results, it is central to identify the relevant state variables: These are in rough Heston models namely (i) the  underlying $(S_t)_{0\leq t\leq T}$ and (ii) the so-called forward variance curve 
\begin{align}\label{eq:ForwardVariance}
(\mathbb{E}[V_{\theta+t}|\mathcal{F}_t])_{\theta\geq 0},
\end{align}
where $V_t$ is the instantaneous variance of the underlying price at
time $t$. In the setting of  \cite{el2018perfect}, the conditional expectation
\eqref{eq:Conditional} above can then be written explicitly as
\begin{align}\label{eq:Conditional2}
C_t=C(T-t, S_t, (\mathbb{E}[V_{\theta+t}|\mathcal{F}_t])_{\theta\geq 0})
\end{align}
where $C$ is some deterministic function.  Indeed it can be observed more broadly, that replicating portfolios in rough volatility models typically contain the \emph{underlying asset} and the  \emph{forward variance curve}. In fact, not only in the case of the rough Heston model but in all affine rough models, a close relation to the forward variance curve can be drawn from the affine structure, as highlighted in Chapter 8.  This gives rise to one of the perspectives presented in Chapter 8, viewing rough affine models 
as forward variance models.  Viens and Zhang in \cite{viens2019martingale} confirm this idea for general Volterra-type stochastic differential equations. In fact, while for some practitioners the idea of using the forward variance curve for hedging (even vanilla options) may come as a surprise, the observation of the importance of the the forward variance curve was already emphasized in \cite{bayer2016pricing} and is also very close in spirit to the approach developed by Bergomi in \cite{bergomi2005smile}. 

\paragraph{Martingale problems and Markovianity}
Viens and Zhang present in \cite{viens2019martingale} a martingale approach, for
general\footnote{This in particular includes the rough Heston model;
  the rough Bergomi model; fractional Ornstein-Uhlenbeck process as
  well as affine Volterra processes.} Volterra stochastic differential
equations. While for affine rough models it has been noted that
martingale problems (connected to hedging problems) are more
convenient thanks to the affine structure, the introduction of a martingale component is also key in the general case in \cite{viens2019martingale} for recovering the flow property, which makes it possible to derive a certain ``Markov'' property for rough models.\\

Pricing and hedging of \emph{volatility options} under rough volatility models has been considered in \cite{horvath2020volatility}, where the special role of the forward variance is re-confirmed as well as martingality considerations revisited. In particular, it is shown that by focusing on the forward variance instead of the instantaneous volatility, one recovers the martingale framework and in particular the classical martingale representation property of option prices.
This makes it possible to compute the hedge ratios, and to show that options can be hedged with a finite number of liquid assets, as in the classical setting. 
To calibrate VIX option smiles via rough volatility we consider extended lognormal models by adding volatility modulation through an independent stochastic factor in the Volterra integral which preserves part of the analytical tractability of the lognormal setting by extending it through an affine structure, which makes it possible to develop approximate option pricing and calibration algorithms based on Fourier transform techniques.\\

In this chapter, we showcase these ideas in relatively transparent and
illustrative settings. We discuss the role of the forward
variance curve to establish (perfect) hedging in rough models, and present a hands-on empirical study illustrating the role of the Hurst parameter (driving the roughness of the paths) on the hedging performance for hedging in VIX options.

\section{A theoretical framework}
The purpose of this section is to illustrate an infinite-dimensional
Markov nature of rough volatility models, which enables us to hedge
options without any ``memory'' of the past.
While fractional Brownian motions have (long or short)
memory properties, we see that the memory is stored in an option market.

\subsection{The model}
Here we consider a 2 factor model; there is a (two-sided) 
2-dimensional standard
Brownian motion $(\bar{W}^1,\bar{W}^2)$ on a probability space
$(\Omega,\mathcal{F},P)$ with
a filtration $\{\mathcal{F}_t\}$ 
being the augmentation of the one generated by the Brownian motion.
We consider a hypothetical option market where call and put options are
traded for all strike prices $K\geq 0$ and maturities $T \geq 0$.
Their prices at time $t \geq 0$ are denoted by $C_t(K,T)$ and $P_t(K,T)$
respectively.
The underlying asset price process of the options is denoted by $S$ and
we suppose
\begin{equation*}
C_t(K,T) = E_Q[(S_T-K)_+|\mathcal{F}_t], \ \ 
P_t(K,T) = E_Q[(K-S_T)_+|\mathcal{F}_t], \ \ S_t = C_t(0,T)
\end{equation*}
for all $K\geq 0$, $T \geq 0$ and $t \geq 0$, where $Q$ is an equivalent measure
to $P$ of which the existence is assumed. 
Here and hereafter we assume risk-free rates are zero
for brevity.

Here we introduce a SABR/Bergomi-type stochastic volatility model
\begin{equation}\label{mod}
 \begin{split}
  & \mathrm{d} S_t = f(S_t)\sqrt{V^t_t}\left[\rho
  \mathrm{d}W^1_t + \sqrt{1-\rho^2}\mathrm{d}W^2_t\right],\\
& \mathrm{d} V^u_t = V^u_t g(u-t)\mathrm{d}W^1_t, \ \ t < u
 \end{split}
\end{equation}
where $(W^1,W^2)$ is a 2-dimensional $\{\mathcal{F}_t\}_{t\geq 0}$-Brownian motion under $Q$,
$f$ and $g$ are deterministic Borel functions on $[0,\infty)$, and $\rho \in
(-1,1)$. 
We assume $g$ is
locally square integrable, so that we have explicit expressions
\begin{equation}\label{exp}
V^u_t = V^u_s \exp\left\{
\int_s^t g(u-v)\mathrm{d}W^1_v - \frac{1}{2}\int_s^t g(u-v)^2\mathrm{d}v
\right\}
\end{equation}
for $0 \leq s \leq t \leq u$.
Note that $E_Q[V^t_t|\mathcal{F}_s] = V^t_s$ for $t \geq s$.
The case $f(s) = s$, $g(u) = \eta u^{H-1/2}$ corresponds to the rough
Bergomi model introduced by \cite{bayer2016pricing}.
A volatility process driven by a fractional Brownian motion can be treated
in this framework. For example, if the log volatility is 
a stationary fractional Ornstein-Uhlenbeck process (see~\cite{BN-BO2011})
\begin{equation*}
 \log v_t = \frac{1}{2}\int_{-\infty}^t g(t-s)\mathrm{d}\bar{W}^1_s, \ \  g(u) = \eta
  u^{H-1/2} - \eta \lambda 
e^{-\lambda u} \int_0^u v^{H-1/2}e^{\lambda v}\mathrm{d}v
\end{equation*}
under $P$ and the volatility risk premium is deterministic, 
then we have (\ref{mod}) and (\ref{exp}) with $f(s) = s$, 
$V^t_u = E_Q[v_t^2|\mathcal{F}_u]$ for $u\leq t$ and a
suitable family $\{V^t_0\}_{t \geq 0}$ of $\mathcal{F}_0$ measurable
random variables (recall that $\bar{W}^1$ is a two-sided Brownian motion). 
We call the curve
\begin{equation*}
\hat{V}_t: \theta \mapsto V_t^{t + \theta}
\end{equation*}
the forward variance curve at time $t$.
\\

\begin{prop}
 The forward variance curve $\{\hat{V}_t\}_{t \geq
0}$ is a Markov process with
state space $C[0,\infty)$.
\end{prop} 
{\it Proof: } By (\ref{exp}), we have for $t \geq s$,
\begin{equation*}
 \hat{V}_t(\theta) = \hat{V}_s(\theta + t-s)
\exp\left\{ \int_s^t g(\theta + t - u)\mathrm{d}W^1_u -
     \frac{1}{2}\int_s^t g(\theta + t- u)^2\mathrm{d}u\right\}.
\end{equation*}
Since the exponential term is independent of $\mathcal{F}_s$, the
result follows.
\hfill////

\begin{cor}
 $(S,\hat{V})$ is a Markov process with state space $[0,\infty)\times
 C[0,\infty)$.
\end{cor}

Now we discuss that $\hat{V}$ is an observable state.
By It\^o's formula, we have
\begin{equation*}
\int_t^{t+\theta} V^u_u \mathrm{d}u =  \int_t^{t+ \theta}
\frac{S_u^2}{f(S_u)^2} \mathrm{d}\langle \log S \rangle_u,
\end{equation*}
which is the payoff of a weighted variance swap.
The fair strike of this swap is
\begin{equation*}
 E_Q\left[
\int_t^{t+ \theta}
\frac{S_u^2}{f(S_u)^2} \mathrm{d}\langle \log S \rangle_u \bigg| \mathcal{F}_t
\right] = \int_t^{t + \theta} E_Q[V^u_u| \mathcal{F}_t]\mathrm{d}u
= \int_t^{t + \theta} V^u_t\mathrm{d}u.
\end{equation*}
Therefore the forward variance curve $\hat{V}$ is the derivative in
$\theta$ of this derivative price.
It is uniquely determined by call and put option prices in a model-free manner as
follows; assume $1/f$ is locally square integrable on $(0,\infty)$ and
let
\begin{equation*}
 h(x) = \int_1^x \int_1^y \frac{2}{f(z)^2}\mathrm{d}z\mathrm{d}y.
\end{equation*}
Then, again by It\^o's formula,
\begin{equation*}
h(S_{t+\theta}) = h(S_t) + \int_t^{t+\theta} h^\prime(S_u) \mathrm{d}S_u 
+ \int_t^{t + \theta} \frac{S_u^2}{f(S_u)^2} \mathrm{d} \langle \log S \rangle_u
\end{equation*}
and by an integration-by-parts formula,
\begin{equation*}
\begin{split}
h(S_{t+\theta}) & = h(S_t) 
+ h^\prime(S_t)(S_{t + \theta} - S_t) \\ &+ 
\int_0^{S_t}(K-S_{t + \theta})_+ h^{\prime\prime}(K)\mathrm{d}K + 
\int_{S_t}^\infty (S_{t+ \theta }-K)_+ h^{\prime\prime}(K)\mathrm{d}K.
\end{split}
\end{equation*}
This means a model-free replication of the weighted variance swap payoff
is
given as a static portfolio of call and put options with weight
$h^{\prime \prime}(K) = 2/f(K)^2$. The replication price is given by
\begin{equation*}
U_t(\theta):= 2 \int_0^{S_t}P_t(K,t + \theta) \frac{\mathrm{d}K}{f(K)^2} + 
2\int_{S_t}^\infty C_t(K,t + \theta)\frac{\mathrm{d}K}{f(K)^2}.
\end{equation*}
Finally we get $\hat{V}_t(\theta)= \frac{\partial}{\partial \theta}
U_t(\theta)$.

Consequently, for a possibly path-dependent functional $F = F(\{S_u\}_{u
\in [t,T]})$, its conditional expectation
$E_Q[F|\mathcal{F}_t]$ is a function of $S_t$ and
$\{\hat{V}_t(\theta)\}_{\theta \geq 0}$, which are observable from the
option market at time $t$.

\subsection{Perfect hedging}
We are considering an infinite dimensional Markov model. But we have only two
factors and so, in light of the martingale representation theorem,
every square integrable payoff is perfectly replicated with a dynamic
portfolio of two traded assets. A natural choice of the two would be
the underlying asset and the weighted variance swap (with a fixed maturity).

As a hedging instrument, the replication portfolio for the weighted variance swap is more
convenient than the weighted variance swap itself because it is a local
martingale. Let
\begin{equation*}
 U^T_t = \int_0^t (h^\prime(S_0)-h^\prime(S_u))\mathrm{d}S_u + 2 \int_0^{S_0}P_t(K,T) \frac{\mathrm{d}K}{f(K)^2} + 
2\int_{S_0}^\infty C_t(K,T)\frac{\mathrm{d}K}{f(K)^2}
\end{equation*}
be the time $t$ value of the replication portfolio with maturity $T$ 
initiated at time $0$.  
We have then
\begin{equation*}
\begin{split}
U^T_t & =  
E_Q\left[  \int_0^{T}
V^u_u\mathrm{d}u \bigg| \mathcal{F}_t \right]
 \\
& = E_Q \left[\int_0^T \left\{V_0^u + \int_0^u V_s^u g(u-s)
    \mathrm{d}W^1_s\right\} \mathrm{d}u\bigg| \mathcal{F}_t\right]\\
& = \int_0^T V_0^u \mathrm{d}u + \int_0^t \mathrm{d}W^1_s \int_s^T V^u_s g(u-s) \mathrm{d}u.
\end{split}
\end{equation*}
Therefore,
\begin{equation}\label{ueq}
 \mathrm{d}U^T_t = \mathcal{D}_gU^T_t\mathrm{d}W^1_t,
\end{equation}
where
\begin{equation*}
  \mathcal{D}_gU^T_t = \int_t^{T}
V^u_t g(u-t)\mathrm{d}u  = 
 \int_t^T\frac{\partial U^{u}_t}{\partial u} g(u-t)\mathrm{d}u.
\end{equation*}
\begin{prop}
For any $F \in L^2(\mathcal{F}_\tau,Q)$, $\tau \in (t,T)$, there exists an adapted process
$(H^S,H^U)$ such that
\begin{equation*}
 F = E_Q[F|\mathcal{F}_t] + \int_t^\tau H^S_v\mathrm{d}S_v + 
\int_t^\tau H^U_v \mathrm{d}U^T_v.
\end{equation*} 
\end{prop}
{\it Proof: }
By the martingale representation theorem, there exists $(H^1,H^2)$ such
that
\begin{equation*}
 F = E_Q[F|\mathcal{F}_t] + \int_t^\tau H^1_v\mathrm{d}W^1_v + 
\int_t^\tau H^2
_v \mathrm{d}W^2_v.
\end{equation*} 
Since
\begin{equation*}
\begin{split}
 & \mathrm{d}W^1_t = \frac{1}{\mathcal{D}_gU^T_t}\mathrm{d}U^T_t, \\
& \mathrm{d}W^2_t = \frac{1}{\sqrt{1-\rho^2}}\left[
\frac{1}{f(S_t)\sqrt{V^t_t}} \mathrm{d}S_t - \rho \mathrm{d}W^1_t
\right]
\end{split}
\end{equation*}
by (\ref{ueq}),
we have the result. \hfill////

\section{Hedging VIX options: empirical analysis}
In this section, we illustrate the advantages of rough volatility
modeling for managing a simple VIX option. We consider three models
for the VIX index: the Black-Scholes model (geometric Brownian
motion), the CIR model, and the rough stochastic volatility model
(where the volatility is the exponential of a fractional Brownian motion). Since we are
interested in hedging
short-term options, we use simplified version of the models without
drift and neglect the effect of the interest rate. As a result, all models have only one parameter to be
estimated (see below).

\begin{figure}
  \centerline{\includegraphics[width=0.6\textwidth]{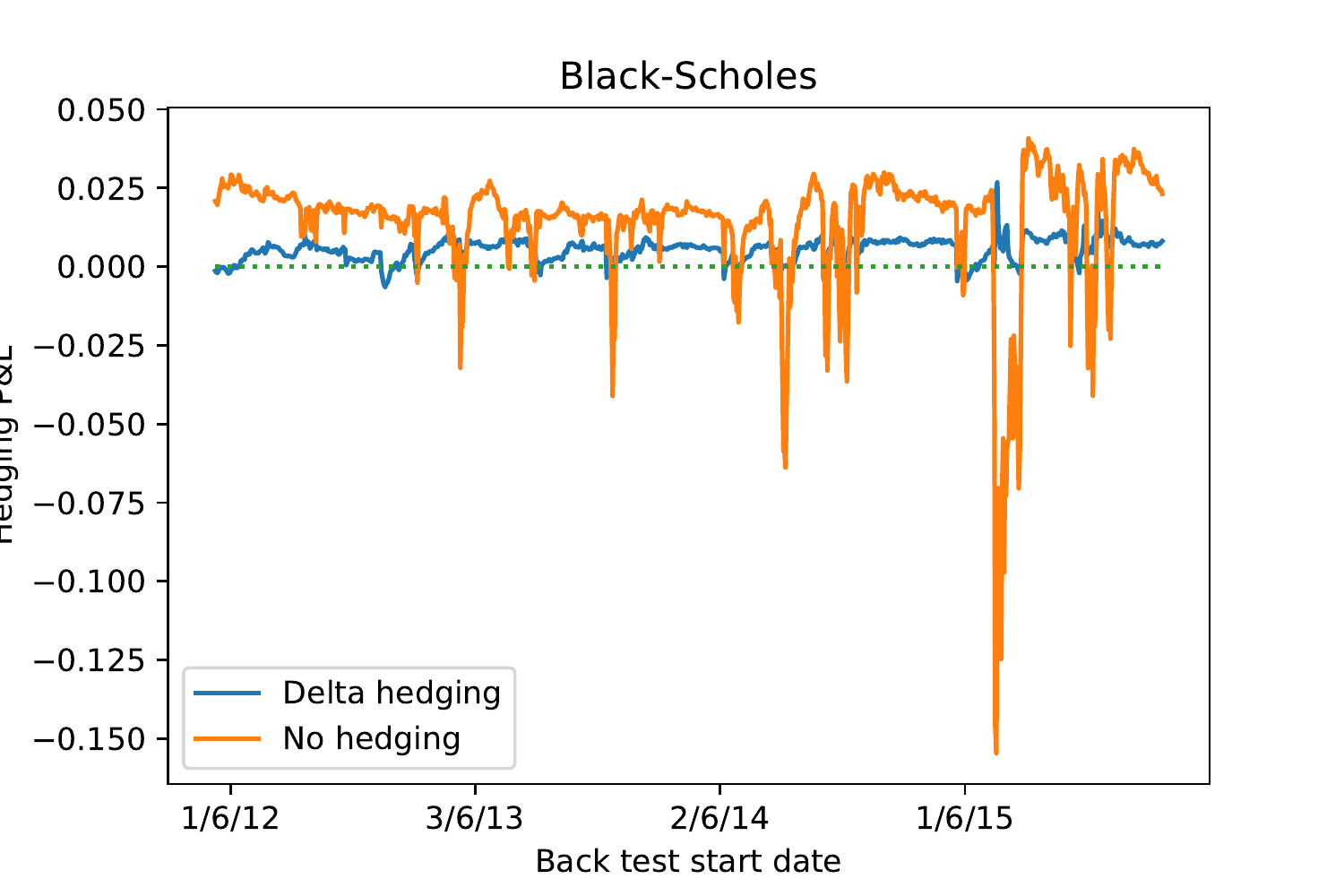}}
  \centerline{\includegraphics[width=0.6\textwidth]{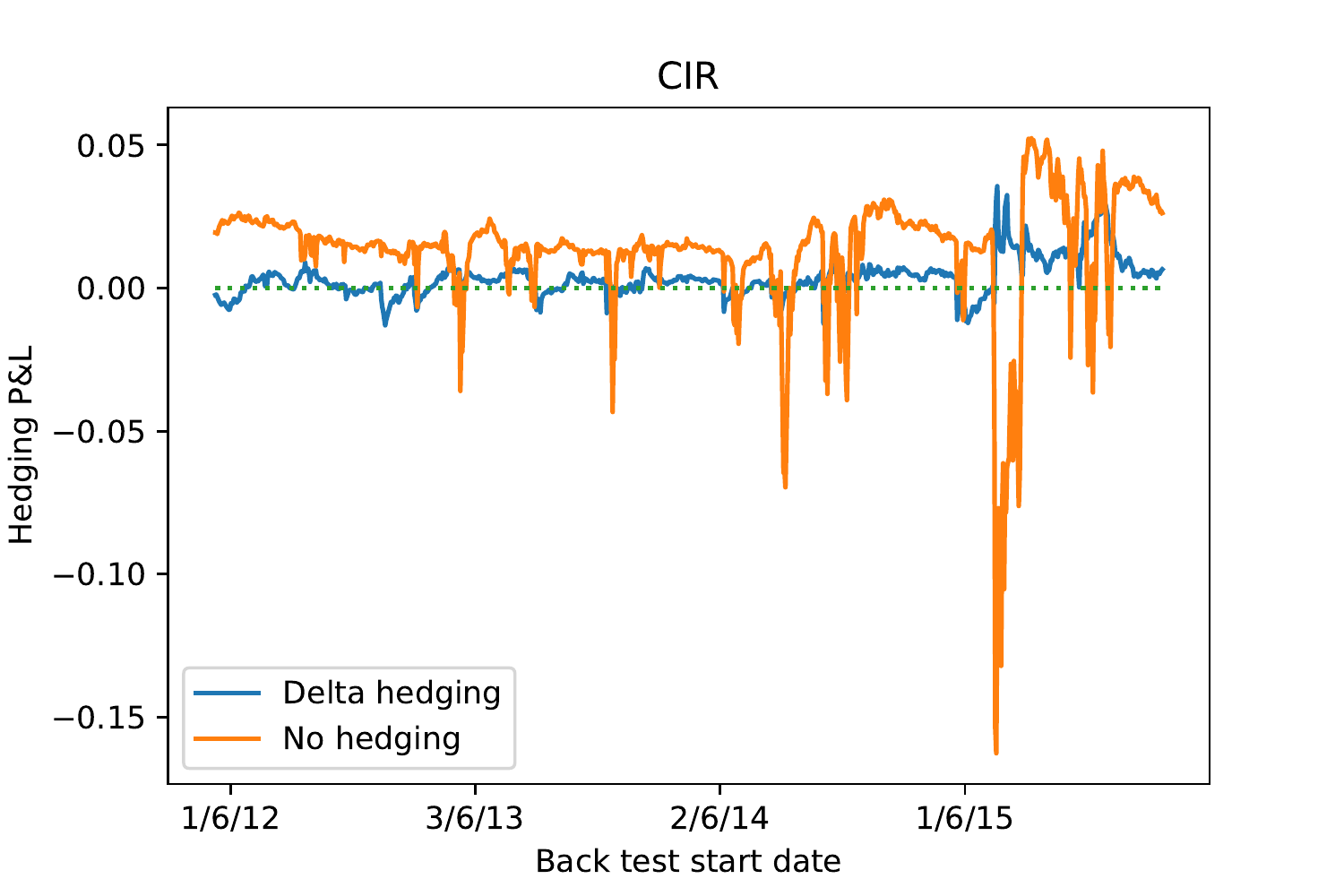}}
  \centerline{\includegraphics[width=0.6\textwidth]{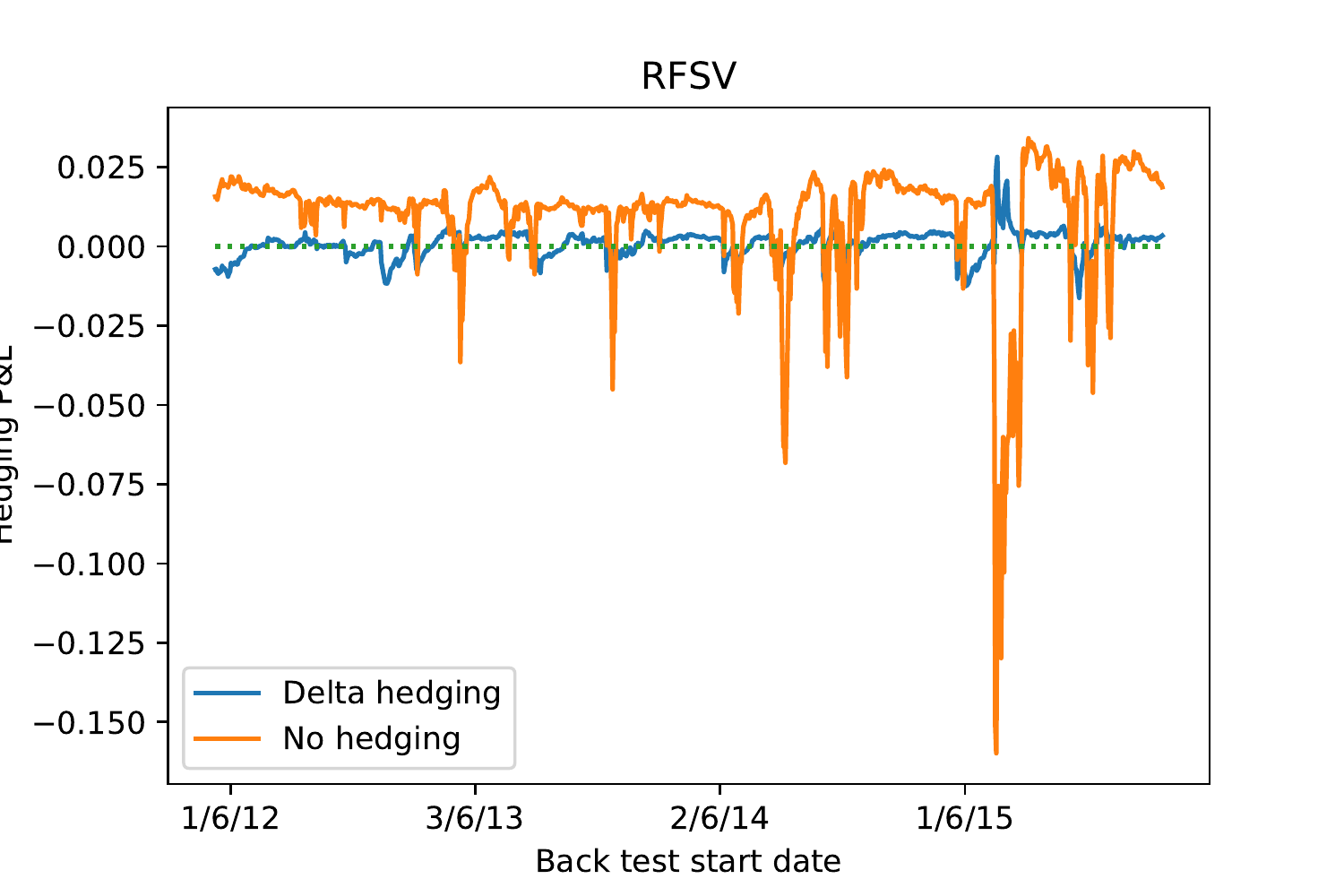}}
  \caption{Back-testing PnL as function of time for the three models
    we study}
  \label{hedge.fig}
\end{figure}

In each model, we perform a series of back-tests of dynamic hedging
of a VIX option with a forward variance swap with the same maturity as
the option and with the duration corresponding to that of the VIX (1
month). In all tests, the hedging portfolio is readjusted daily using
the closing prices of the hedging instruments. The test is performed 1000 times, starting on each working day $t$
between Jan 10, 2012 and Apr 29, 2016. The back-test is organized as follows:
\begin{itemize}
  \item The parameter is estimated on the 88-day period preceding day
    $t$.
   \item The initial value of the hedging portfolio is initialized
     with the ATM VIX option price with maturity 1.5 months computed
     within the model, and the quantity of the hedging asset in the
     portfolio is initialized with the corresponding model-based hedge
     ratio.
     \item For 29 working days following day $t$, each day the
       portfolio value is readjusted following the change in the value
       of the hedging asset, and the hedge ratio is recomputed.
      \item At the end of the 29 working day period, the P\&L of the
        hedging portfolio is recorded, and the no-hedge P\&L is
        recorded as the difference between the option price at the
        beginning and at the end of the period.  
      \end{itemize}
The back test uses synthetic forward variance curve data, computed
from the historical prices of S\&P index options, downloaded from the
WRDS database.       
The detailed description of models and hedging procedures is given below. Table \ref{hedge.tab} presents the main
results of the back test. We see that while the Black-Scholes and CIR
benchmarks appear to have similar performance, the RFSV model
exhibits a much lower bias, and a RMSE which is 27\% lower than the
other two models. Figure \ref{hedge.fig} plots the back-test PnL evolution as
function of the starting date of the back test. The consistently low
bias of the strategy based on the RFSV model is clearly
visible here. 

\begin{table}
  \begin{tabular}{lllllll}
    &\multicolumn{2}{c}{Black-Scholes}&\multicolumn{2}{c}{CIR}&\multicolumn{2}{c}{RFSV}\\
    &No hedge & Hedge & No hedge& Hedge&No hedge & Hedge
    \\
    Mean & 0.01445 &  0.005336 & 0.01363 & 0.003399 & 0.009919 &0.0006345 \\
Std.~dev. &0.01896 & 0.003555& 0.02069 & 0.006506 &0.01880&0.004141\\
RMSE & 0.02384 & 0.006412  & 0.02478 &0.007341 & 0.02125 & 0.004190 \\
 Red.~ factor &    3.7176 & &        3.7176 &    &      5.0724 &   
  \end{tabular}
  \label{hedge.tab}
  \caption{Empirical Performance of hedging strategies based on
    different models}
\end{table}  

\paragraph{Black-Scholes model}

Let $\text{VIX}_t$ be the VIX index at time $t$, and let $F^T_t$ be the T-forward variance swap at time $t$, which refers to the same period as the VIX index, that is, $F^T_t = E_Q[\text{VIX}^2_T|\mathcal F_t]$. Assume that the VIX index follows the log-normal dynamics $\text{VIX}_t = e^{X_t}$, where $X$ is an OU process $\mathrm{d}X_t = \kappa(\theta-X_t) \mathrm{d}t + \gamma \mathrm{d}W_t$ under the risk-neutral probability. Then, forward variance swap has dynamics
$$
\mathrm{d}F^T_t = 2F^T_t e^{-\kappa(T-t)} \gamma \mathrm{d}W_t. 
$$
When close to term, the exponential factor can be neglected and we obtain the simple Black-Scholes dynamics. On the other hand, $\gamma$ may be estimated from the volatility of VIX:
$$
\mathrm{d}\langle \text{VIX}\rangle_t = \text{VIX}_t^2 \gamma^2 \mathrm{d}t.
$$
We are hedging a VIX option with pay-off 
$$
\left(\text{VIX}_T-K\right)_+.
$$
Introducing the VIX future
$$
\text{VIX}^T_t =  E_Q[\text{VIX}_T|\mathcal F_t],
$$
neglecting the interest rate, the option price is given by 
$$
p(t,\text{VIX}^T_t) =\text{VIX}^T_t N({d^1_t}) - K N({d^2_t}),\quad {d_t^{1,2}} = \frac{\log\frac{\text{VIX}^T_t}{K}\pm \frac{\gamma^2(T-t)}{2}}{\gamma\sqrt{T-t}},
$$
or in other words,
$$
p(t,F^T_t) = \sqrt{F^T_t} e^{-\frac{\gamma^2}{2}(T-t)} N({d^1_t}) - K N({d^2_t}),\quad {d_t^{1,2}} = \frac{\log\frac{\sqrt{F^T_t} e^{-\frac{\gamma^2}{2}(T-t)}}{K}\pm \frac{\gamma^2(T-t)}{2}}{\gamma\sqrt{T-t}},
$$
and the hedge ratio is
$\frac{N({d^1_t})}{2\sqrt{F^T_t}}e^{-\frac{\gamma^2}{2}(T-t)}$.
\paragraph{CIR model}
Assume that the VIX index follows the square root dynamics:
$$
\mathrm{d}\text{VIX}^2_t = \kappa(\theta-\text{VIX}^2_t) \mathrm{d}t + \gamma \text{VIX}_t \mathrm{d}W_t.
$$
Since we are hedging short maturity options and cannot estimate $\kappa$ and $\theta$ under the risk-neutral measure anyway, we assume that $\kappa=0$ so that 
$$
\mathrm{d}\text{VIX}^2_t =  \gamma \text{VIX}_t \mathrm{d}W_t.
$$
The forward variance swap is then given by
$$
F^T_t = E_Q[\text{VIX}_T^2 | \mathcal F_t] = \text{VIX}^2_t,
$$
and follows the dynamics
$$
\mathrm{d}F^T_t = \gamma\sqrt{F^T_t} \mathrm{d}W_t
$$

We are hedging a VIX option with pay-off
$$
\left(\text{VIX}_T - K\right)_+
$$
with a forward variance swap.
The price of the VIX option is given by
\begin{align*}
P(t,F^T_t) & = \int_{K^2}^\infty (\sqrt{x}-K) p_{T-t}(F^T_t,x) \mathrm{d}x,
\end{align*}
where $p_T(v_0,x)$ is the density of the CIR process at time $T$ with
the starting value $v_0$. The parameter $\gamma$ may be estimated by
observing that $\langle {\text{VIX}}\rangle_t = \frac{\gamma^2}{4}t$. 

\paragraph{Rough fractional stochastic volatility}
Assume now that the VIX index is given by    
$$
\text{VIX}_t = C e^{X_t},
$$
where {$C>0$ is a constant and} $X$ is a centered Gaussian process under the risk-neutral
probability.
For all $s\geq 0$, let $\mathcal F^0_s := \sigma(X_r,r\leq s)$, 
and $\mathcal F_s := \cap_{s<t} \mathcal F^0_t$. 
The interest rate is taken to be zero. 
Fix a time horizon $T$, let $Z_t(T) := E_Q[X_T|\mathcal F_t]$,
so that $(Z_t(T))_{t\geq 0}$ is a Gaussian martingale and thus a process with independent increments, completely characterised by the function 
$$
c^T(t) := E_Q[Z_t(T)^2] = E_Q[E_Q[X_T|\mathcal F_t]^2]. 
$$
If we assume in addition that $c^T(\cdot)$ is continuous then $(Z_t(T))_{t\geq 0}$ is almost surely continuous.
Using the total variance formula, the forward variance swap can be characterised as
$$
F^T_t := E_Q[\text{VIX}^2_T|\mathcal F_t] = {C^2}E_Q[e^{2X_T}|\mathcal F_t]
 = {C^2}e^{2E_Q[X_T|\mathcal F_t]+2 \text{Var}[X_T|\mathcal F_t]}
 =  {C^2} e^{2(Z_t(T) + c^T(T) - c^T(t))}.
$$
The time-$t$ price of a Call on the VIX
is given by
$P_{t}:=E_Q[(\text{VIX}_T-K)^+ |\mathcal F_t]$.
Note that the VIX future is a continuous lognormal martingale with 
$E_Q[\text{VIX}_T|\mathcal F_t] = \text{VIX}^T_t$
and, by the total variance formula,
$$
\text{Var}[\log \text{VIX}_t|\mathcal F_t] = \text{Var}[X_{T}|\mathcal F_t]= c^T(T) - c^T(t).
$$
In other words, 
$$P_t = \text{VIX}^T_t N(d^1_t) - K N(d^2_t),\quad
d^{1,2}_t = \frac{\log\frac{\text{VIX}^T_t}{K}\pm {\frac{1}{2}}(c^T(T) - c^T(t))}{\sqrt{c^T(T) - c^T(t)}}.
$$
Applying It\^o's formula 
and keeping in mind the martingale property of the option price, we obtain
$$
\mathrm{d}P_t = N(d^1_t) \mathrm{d}\text{VIX}^T_t.
$$
In terms of forward variance swap, we then have: 
$$
P_t = \sqrt{F^T_t} e^{-\frac{1}{2}(c^T(T)-c^T(t))} N(d^1_t) - K N(d^2_t),\quad
d^{1,2}_t =
\frac{\log\frac{\sqrt{F^T_t}e^{-\frac{1}{2}(c^T(T)-c^T(t))}}{K}\pm
{\frac{1}{2}}(c^T(T) - c^T(t))}{\sqrt{c^T(T) - c^T(t)}},
$$
and the option price dynamics takes the following form:
$$
\mathrm{d}P_t = \frac{N(d^1_t) e^{-\frac{1}{2}(c^T(T)-c^T(t))}}{2\sqrt{F^T_t}} \mathrm{d}F^T_t
$$
Assuming that 
$$
X_t = \sigma W^H_t,
$$
where $W$ is the fractional Brownian motion with the Hurst parameter $H$, 
we get, after some computations using the Mandelbrot-Van Ness representation:
\begin{align*}
c^T(t) &= \frac{\sigma^2}{\Gamma^2(H+1/2)}\int_0^\infty
         \left[(T+s)^{H-1/2} - s^{H-1/2}\right]^2\mathrm{d}s +
         \frac{\sigma^2}{\Gamma^2(H+1/2)}\int_0^t (T-s)^{2H-1}\mathrm{d}s\\  &= f(T) - \frac{\sigma^2 (T-t)^{2H}}{2H\Gamma^2(H+1/2)}
\end{align*}
for some function $f(T)$, which cancels out in the difference, so that 
$$
c^T(T)-c^T(t) = \frac{\sigma^2 (T-t)^{2H}}{2H\Gamma^2(H+1/2)}.
$$

\begin{figure}
  \centerline{\includegraphics[width=0.7\textwidth]{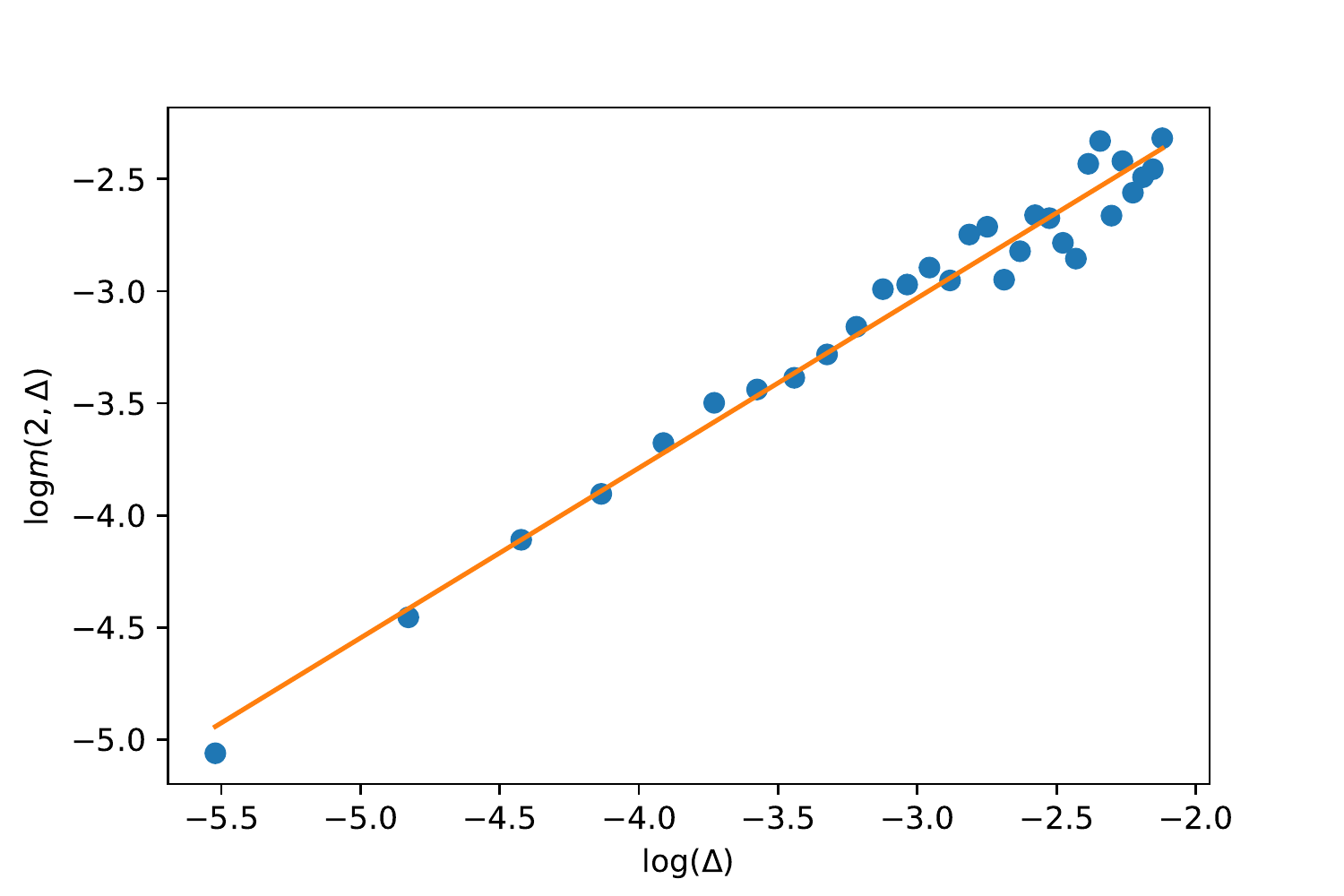}}
  \caption{Estimation of the Hurst parameter}
  \label{hestim.fig}
\end{figure}

Contrary to the previous two models, this one formally has two parameters to be
estimated: $\sigma$ and $H$. To estimate the Hurst parameter, following \cite{gatheral2018volatility}, we define
$$
m(q,\Delta) = \frac{1}{\lfloor T/\Delta\rfloor} \sum_{k=1}^{\lfloor T/\Delta\rfloor} |\log(\text{VIX}_{k\Delta})-\log(\text{VIX}_{(k-1)\Delta})|^q,
$$
and estimate $H$ from the half slope of $m(2,\Delta)$ as function of
$\Delta$ in the log-log coordinates (see Figure
\ref{hestim.fig}, where $\Delta$ varies from 1 to 30 days). Since this
procedure requires a relatively long dataset to be precise, we perform
it only once, on the VIX index time series from April 17, 2001 to
April 16, 2021. This gives an estimated Hurst parameter value of
$0.377$, and the procedure is quite stable: when estimating on the first 10 years of
the  dataset, one obtains $0.380$ and on the last 10 years one obtains
$0.379$.

These estimated values of the Hurst index are much higher than the values found by
\cite{gatheral2018volatility} and many other authors using
the daily time series of realized volatility (typically between 0.1
and 0.15). However, the VIX index is constructed from prices of
one-month options on the S\&P index, and using the implied volatility of one-month options
as proxy for volatility, \cite{livieri2018rough} find a value of
$H=0.32$, which is much closer to our result. The relatively high
value of the Hurst index we find can thus be explained by the
averaging effects associated with computing option prices. 

In view of the stability of the Hurst index estimation, we fix the value $H=0.377$ for all
tests, rather than estimating it before each back-test. Note that the
hedging performance of the model remains very similar for $H\in
[0.37,0.39]$. This leaves us with a single parameter, $\sigma$, to
estimate before each back test, which is estimated by
$$
\hat \sigma = \sqrt{\frac{m(2,\Delta)}{\Delta^{2H}}}. 
$$

\begin{figure}
  \centerline{\includegraphics[width=0.7\textwidth]{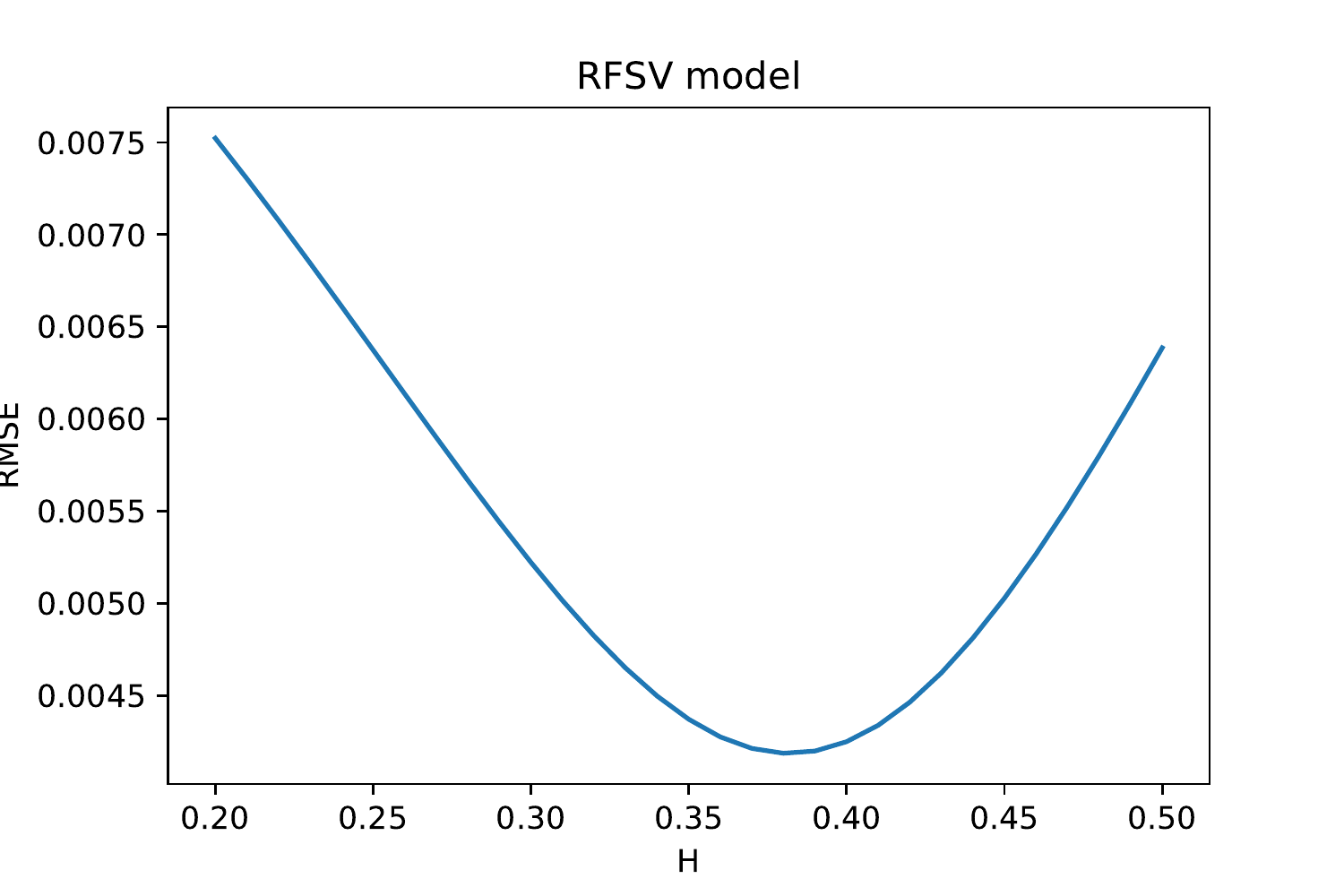}}
  \caption{Dependence of the hedging RMSE on the Hurst parameter value.}
  \label{hvar.fig}
\end{figure}

To further illustrate the dependence of the hedging performance on the
value of the Hurst index and the importance of using a 'rough
volatility' specification, we performed the same test for
$H$ values ranging between $0.2$ and $0.5$, with step of $0.01$, where
the value $H=0.5$ corresponds to the Black-Scholes benchmark. Figure
\ref{hvar.fig} shows the dependence of the hedging RMSE on the value
of the Hurst parameter with the minimum attained around $H=0.38$. 

\end{document}